\documentclass[conference]{IEEEtran}

\makeatletter
\def\ps@headings{%
\def\@oddhead{\mbox{}\scriptsize\rightmark \hfil \thepage}%
\def\@evenhead{\scriptsize\thepage \hfil \leftmark\mbox{}}%
\def\@oddfoot{}%
\def\@evenfoot{}}
\makeatother
\usepackage{algorithm}
\usepackage{algorithmicx}
\usepackage{algpseudocode}
\usepackage{amsmath}
\usepackage{graphicx}
\usepackage{subfigure}
\usepackage{times}
\usepackage[hyphens]{url}
\usepackage{color}
\usepackage{multirow}
\usepackage{textcomp}
\usepackage{threeparttable}
\usepackage{supertabular}
\usepackage{amssymb}
\usepackage{amsfonts}
\usepackage{booktabs}
\usepackage{epstopdf}
\usepackage{xspace}
\usepackage{enumitem}
\usepackage{cite}

\newcommand{\BotShape}{{\sc BotShape}\xspace}
\newcommand{\Cresci}{{\sc Cresci2017}\xspace}
\newcommand{\Twitter}{{\sc Twitter}\xspace}
\renewcommand{\paragraph}[1]{\smallskip\noindent {\bf #1}}

\begin{document}

\title{BotShape: A Novel Social Bots Detection Approach via Behavioral Patterns}

\author{Jun Wu$^1$, Xuesong Ye$^2$, Chengjie Mou$^2$\\
$^1$Georgia Institute of Technology
\\
$^2$Trine University}

\author{
  \IEEEauthorblockN{Jun Wu\IEEEauthorrefmark{1}, Xuesong Ye\IEEEauthorrefmark{2}, and Chengjie Mou\IEEEauthorrefmark{2}}
  \IEEEauthorblockA{\IEEEauthorrefmark{1}Georgia Institute of Technology\\ Email: jwu772@gatech.edu}
  \IEEEauthorblockA{\IEEEauthorrefmark{2}Trine University\\ Email: \{xye221,cmou22\}@my.trine.edu}
}

\maketitle

\begin{abstract}
An essential topic in online social network security is how to accurately detect bot accounts and relieve their harmful impacts (e.g., misinformation, rumor, and spam) on genuine users. Based on a real-world data set, we construct behavioral sequences from raw event logs. After extracting critical characteristics from behavioral time series, we observe differences between bots and genuine users and similar patterns among bot accounts. We present a novel social bot detection system \BotShape, to automatically catch behavioral sequences and characteristics as features for classifiers to detect bots. We evaluate the detection performance of our system in ground-truth instances, showing an average accuracy of 98.52\% and an average f1-score of 96.65\% on various types of classifiers. After comparing it with other research, we conclude that \BotShape is a novel approach to profiling an account, which could improve performance for most methods by providing significant behavioral features.
\end{abstract}

\begin{IEEEkeywords}
Social bots detection, Behavioral features mining, Machine learning
\end{IEEEkeywords}

\section{Introduction}
\label{sec:intro}

\subsection{Background and Motivation}
It had become common sense that bot accounts flooded almost every popular social network platform. Many research works have realized these issues and made solutions like Twitter \cite{cresci2017paradigm}, Facebook \cite{santia2019detecting}, and Reddit \cite{hurtado2019bot}. Along with the breaking out social bot population, user-experience of normal users deteriorate quickly due to malicious bot direct or indirect disturb, e.g., fake news \cite{heidari2021bert}, spam \cite{rodrigues2022real}, rumor \cite{huang2022social}, and misinformation\cite{himelein2021bots}. Furthermore, it will finally push platforms to improve a large amount of maintenance and risk control costs or otherwise suffer from user loss. Therefore, designing highly accurate and easily deployed bot detection systems is substantial for research and industry.

\subsection{Bot Detection Approaches}
\label{subsec:related}
Recent social bot detection works could be categorized into three strategies: account-based, content-based, and graph-based. 

\textbf{Account-based} approach mainly focuses on mining risk signatures as features from account meta-data and statistical indicators for machine learning algorithms to do classification. Kai-Cheng {\em et al.} \cite{yang2020scalable} extracted two types of features, including raw data like follower count and derived features like follower growth rate and the length of the screen name. Saleh {\em et al.} \cite{ahmad2021spam} summarized a bunch of features according to account information like age, length of the name, and count of followers and then used a support vector machine to separate bots from genuine users. Hrushikesh {\em et al.} \cite{shukla2021enhanced} ensemble the prediction results of three machine learning models to improve the classifying accuracy after a comprehensive account features collection, e.g., location, profile image, and daily average tweet count, and so on. Maryam {\em et al.} \cite{heidari2022online} utilized GloVe \cite{pennington2014glove} to create account embedding based on age, gender, education, and personality from meta-data, which firstly proposes the user embedding concept.

\textbf{Content-based} method systems shift the spotlight to public texts posted by users, like tweets,  and apply natural language processing techniques to mine risk words or embedding semantics. The primary assumption of those works is that content posted by bots tends to uncover their fraud intentions,  and various intentions are clusters in high-dimensional language embedding space. Anisha {\em et al.} \cite{rodrigues2022real} applies the TF-IDF and Bag of Words technique to generate text features for the downstream machine learning model to train and predict. BGSRD \cite{guo2021social} combined Bert \cite{devlin2018bert} and GCN(Graph Convolutional Networks) \cite{kipf2016semi} algorithm to jointly learn representation from multiple historical tweets of each account for bot classification tasks. DeepSBD \cite{fazil2021deepsbd} learned text representation based on historical tweets and mixed content embedding with account features via joint representing. Maryam {\em et al.} \cite{heidari2021bert} also uses BERT to generate text embedding from tweets, showing a significant performance in detecting fake news about the COVID-19 topic. 

\textbf{Graph-based} approach gets popular in the bot detection domain after the rapid evolution of the graph representing techniques. After 2016, newly proposed algorithms GraphSAGE \cite{hamilton2017inductive} and GCN \cite{kipf2016semi} perform significantly among various network relationships in the real world, like social networks, online shopping, and citation map. The graph-based approach reuses useful account-based and content-based features by transforming them into node embedding. Moreover, it explores the optimal network topology to share and transfer node information among neighbors, outperforming traditional methods. Some recent works show a start-of-art accuracy by designing appropriate graph structures and node features. Seyed {\em et al.} \cite{ali2019detect} firstly apply graph convolutional neural networks to learn one node's representation based on account features of itself and its neighbors. Shangbin {\em et al.} \cite{feng2021botrgcn} applied GCN algorithm on the user following relationship graph, then represented raw node features including user profile, categorical and numerical data of account activity. Shangbin {\em et al.} \cite{feng2022heterogeneity} constructs two kinds of heterogeneity structures, including relation and influence, leveraging the topology to identify the difference between genuine users and social bots.

\subsection{Existing Problems}
\label{subsec:prob}
Most \emph{account-based works} focus on evaluating the effects of different machine learning models without a deep digging into the behavior patterns of bots. 
\textbf{The first problem} is that most works tend to input coarse-grain statistical indicators as features into models without controlling variables. For example, the count of followers is an essential feature in many bot detection models. However, the count of followers of one account will naturally increase and accumulate day-to-day after registering. Engineers usually calculate the indicators on the day of building models. However, the registering dates of accounts are different. As a result, these calibrated features become noisy data to prevent the classifier from making an accurate prediction. In short,  adequate log data is not utilized sufficiently in this way. 
\textbf{The second problem} is that most works consider bots isolated rather than gangs of attackers. Consequently, little research is focusing on discovering behavior similarity among bot accounts.
In \emph{graph-based methods}, accounts will exchange node information with neighbors. However, most focus on finding a better network topology rather than exploring behavioral node features.

\subsection{Our Contributions} 
To summarize, this paper makes the following contributions:
\begin{itemize}
\item We first propose a novel type of feature to profile a social account's behavioral pattern, supported by a solid measurement work to observe distribution divergence between genuine users and bot accounts. Based on behavioral sequences, we adopt proper algorithms to catch their significant patterns for prediction. 

\item We design a bot detection system \BotShape integrating an automatic behavioral feature generation process and implementing a complete pipeline log processing, feature engineering, and prediction. Input parameters are simplified into a minimal range to improve the automation degree of the whole process. 

\item \BotShape performs high accuracy, with an average accuracy of 98.52\% and an average f1-score of 96.65\% in evaluating experiments across four classifiers and three ground truths. It presents a steady performance improvement compared to \emph{account-based features} and also shows the crucial role of extracting\emph{behavioral patterns}. \BotShape is easy to combine with other bot detection systems by providing compelling behavioral features. For example, \emph{graph-based} approach could use behavioral features as the node features and 
spread pattern information into the whole network.

\end{itemize}

\section{Dataset}
\label{sec:dataset}

\begin{table*}[h]
\centering
\caption{Key attributes of account and tweeting data in \Cresci data set}
\label{tab:attributes}
\vspace{-6pt}
\renewcommand{\arraystretch}{1.1}
\small
\begin{tabular}{|l|p{4.5in}|}
\hline
{\bf Attribute} & {\bf Descriptions}\\
\hline
User ID & It refers to the unique identifier for each tweet account. \\
\hline
Created At & It referred to the time stamp at the time of the account registration.  \\
\hline
Account Information & There are several essential attributes, such as the nickname of an account, the personal introduction, whether the avatar is the default, and the common location.
\\
\hline
Interaction (user-level) & They refer to the number of accounts that interact with an account, potentially reflecting the social vitality of an account, such as the count of followers, friends, and favorites. \\
\hline
Tweet ID & It refers to the unique identifier for each tweet post. It has a unique id of its author. \\
\hline
Tweet Created At & It is the time stamp of the posting time of a work like a tweet.\\
\hline
Post Content & It refers to the raw text or voice of a post.\\
\hline
Interaction (tweet-level) & They are statistics of re-tweeting and replying to a post, reflecting its popularity. \\
\hline
\end{tabular}
\end{table*}
Cresci {\em et al.} \cite{cresci2017paradigm} published a data set of Twitter (called \Cresci) of four types of accounts, including genuine users, social bots, traditional users, and fake followers. It collected user information and event logs (mainly tweeting records with timestamps, identifiers, and text) in the real world. The total number of social bots is 4912, consisting of three small data sets collected at different periods (from the easiest tweet publish time to the last one): (i) from March 17, 2009, to May 26, 2014 (ii) from September 9, 2008, to March 22, 2014 (iii) from September 14, 2008, to April 11, 2014. The data group of genuine users begins on January 22, 2007, and ends on April 20, 2015, with 3474 accounts in total (only 1083 of them have tweeting logs). There are two extra spam account data sets with complete tweeting logs: (i) traditional spam bots: from July 4, 2007, to March 8, 2010, with 1000 accounts that have tweeting logs (ii) fake followers: from December 7, 2007, to April 30, 2013, with 3351 accounts. The collected tweets come across a very long period, sufficient to observe accounts' early and long-term behavior. \Cresci supports abundant real-world user and tweet data for analysis and modeling. Based on domain knowledge, we highlight important ones and summarize them in Table~\ref{tab:attributes}.

\section{Measurement}
\label{sec:measure}
In our work, we commit to solving the two problems discussed in sub-section~\ref{subsec:prob}, through refined behavior indicators and solid measurement results. We discover characteristics of accounts' behavior patterns and uncover hidden correlations between and show how different they are between bot accounts and genuine users. All measurement results support a proper architecture of \BotShape in Section~\ref{sec:design}, including feature extraction and prediction model.

We split measurement work into two parts: (i) The first part shows how to construct calibrated indicators without a \emph{life-cycle} inconsistency problem. Then we analyze how the calibrated indicators change over time on a real-world data set. Furthermore, we also compare the fluctuating pattern between bot accounts and genuine users. (ii) The second part digs into tweet count analysis from a time series perspective. We observe the busy and idle time in the one-day and one-week range. Then we observe the clusters of behavioral time series to verify the assumption that bots have more synchronized behaviors than genuine users because of the centralized control of the dark industry.

\subsection{Account Life-cycle and Cleaning Inactive ones}
\subsubsection{\textbf{Account life-cycle and event logs}} Each social network account has its life-cycle, starting from the registration time to the current system time. One account did many actions in its life-cycle, e.g., following, liking, posting,  restored as \textbf{event logs} by the system. Behavioral logs accumulate day-by-day as time goes on; as a result, indicators like the count of tweets increase and change. Registration time-stamp is a vital starting point to calibrate event logs. We construct a new metric $Indicator_{dur}$  using $dur$ as a variable that refers to the duration from the registration to the current date-time. For fairness, we only compare indicators among accounts with the same parameter $dur$.

With complete tweeting records, we could calculate multiple behavior indicators with different $dur$ parameters at various positions in one account's life-cycle. For a better observation, we select multiple $dur$ variables with the same time interval. The definition of $dur$ in the measurement is the count of days from the registration to the analysis date-time. For example, if one account registered on May 1, 2014, and the virtual analysis happened on May 3, 2014, the $dur$ equals two days.

\subsubsection{\textbf{Inactive accounts}} We calculate $Indicator_{dur}$ for each account after removing the inactive ones. We define \textbf{active account} as one having at least one tweet in the first month (30 days) after registering. Inactive accounts, most are audiences, present less behavior. They occupy a large proportion of the data set. We remove them to focus on informative and active behavior data rather than letting inactive accounts average and weaken important distribution regularity. For a fair comparison, we remove \textbf{inactive accounts} from the bot accounts and normal users simultaneously.

\begin{figure}[h]
\centering
\begin{tabular}{c}
\includegraphics[width=3.3in]{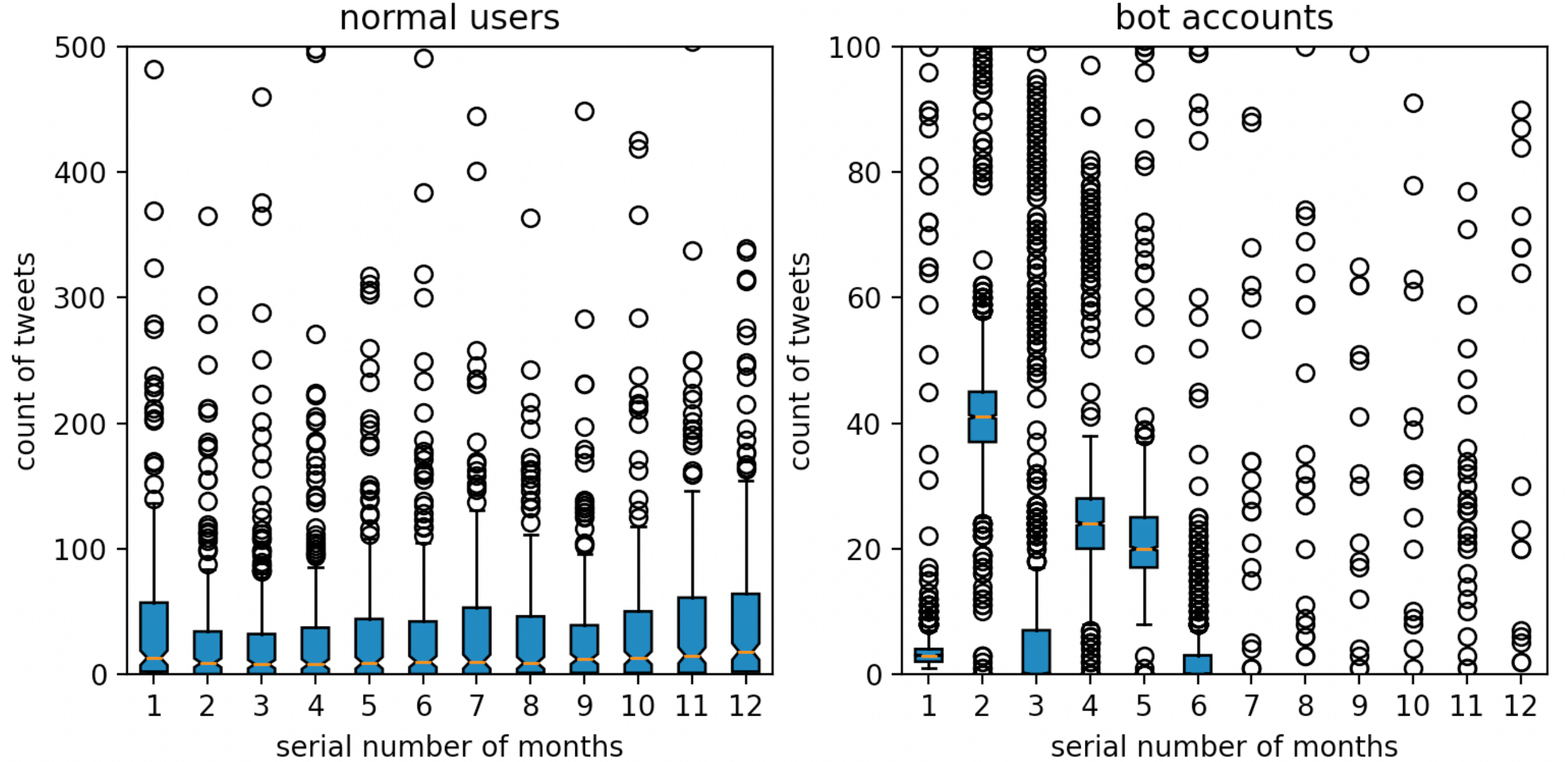}
\vspace{-3pt}\\
\mbox{\small (a) Monthly count of tweets}
\vspace{3pt}\\
\includegraphics[width=3.3in]{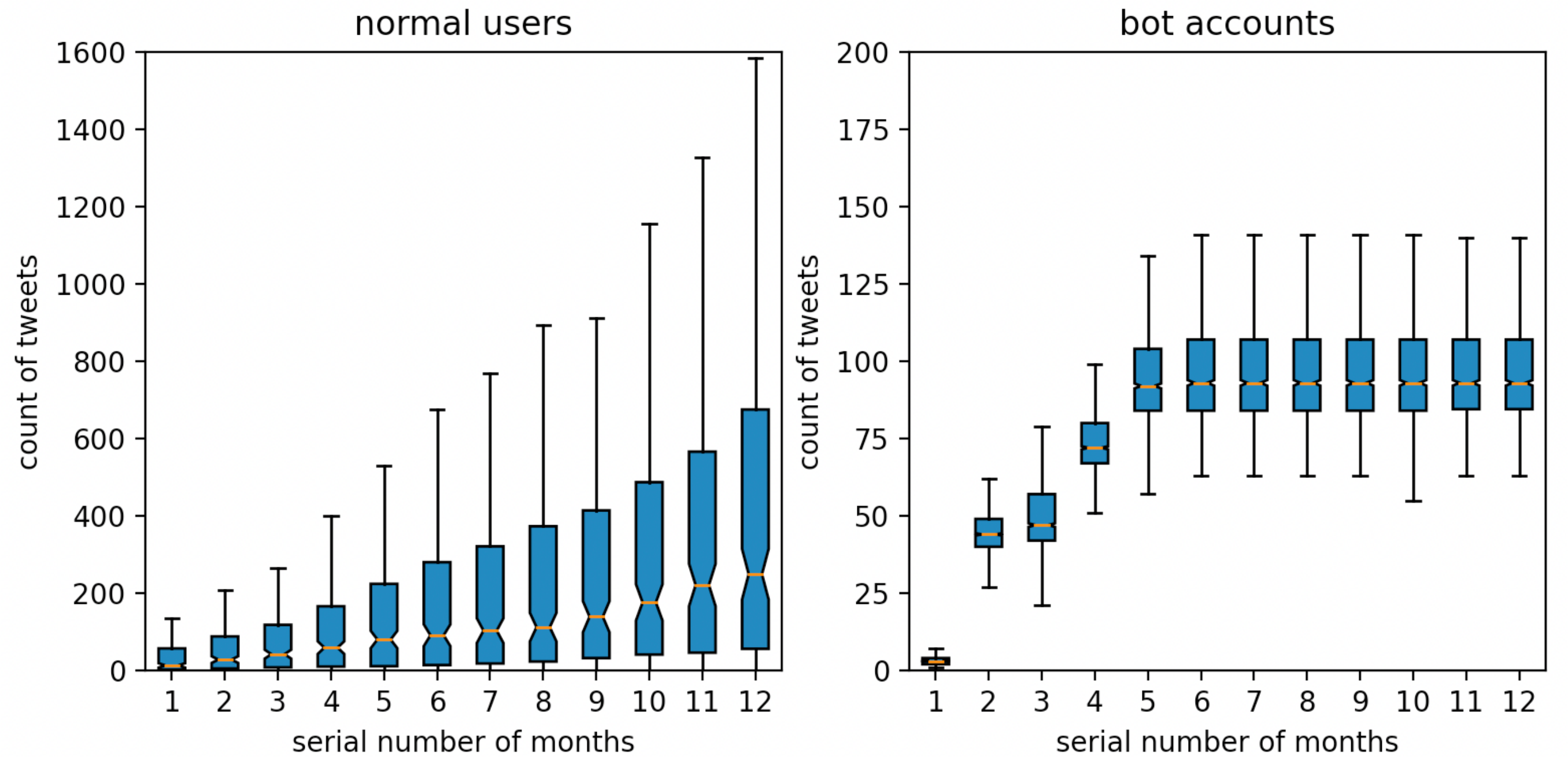}
\vspace{-3pt}\\
\mbox{\small (b) Monthly accumulated count of tweets}
\end{tabular}
\caption{Boxplots of tweeting counts: bot  accounts vs normal users}
\label{fig:monthly_tweets}
\end{figure}

\begin{figure}[h]
\centering
\begin{tabular}{c}
\includegraphics[width=3.3in]{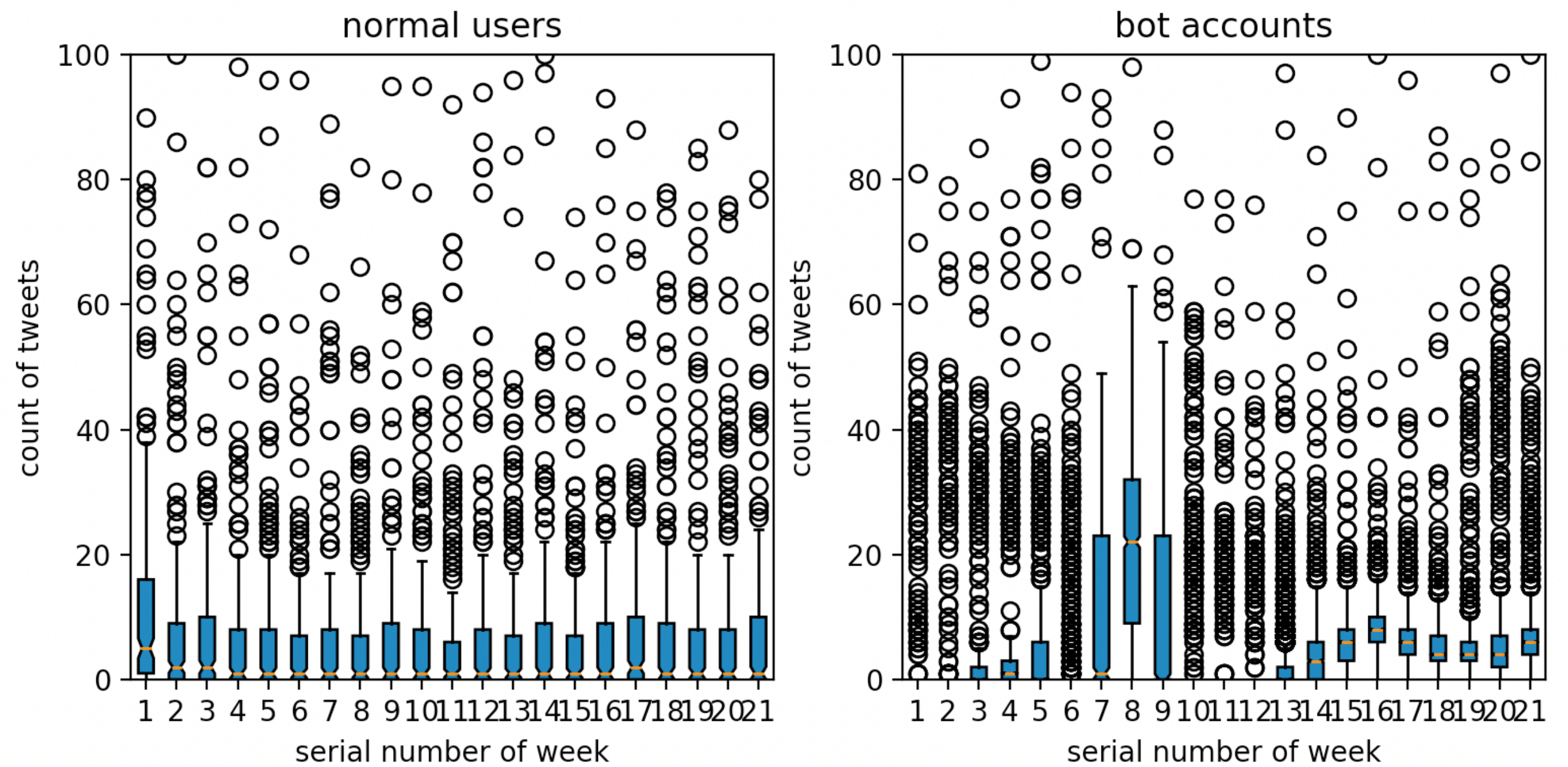}
\vspace{-3pt}\\
\mbox{\small (a) Weekly count of tweets}
\vspace{3pt}\\
\includegraphics[width=3.3in]{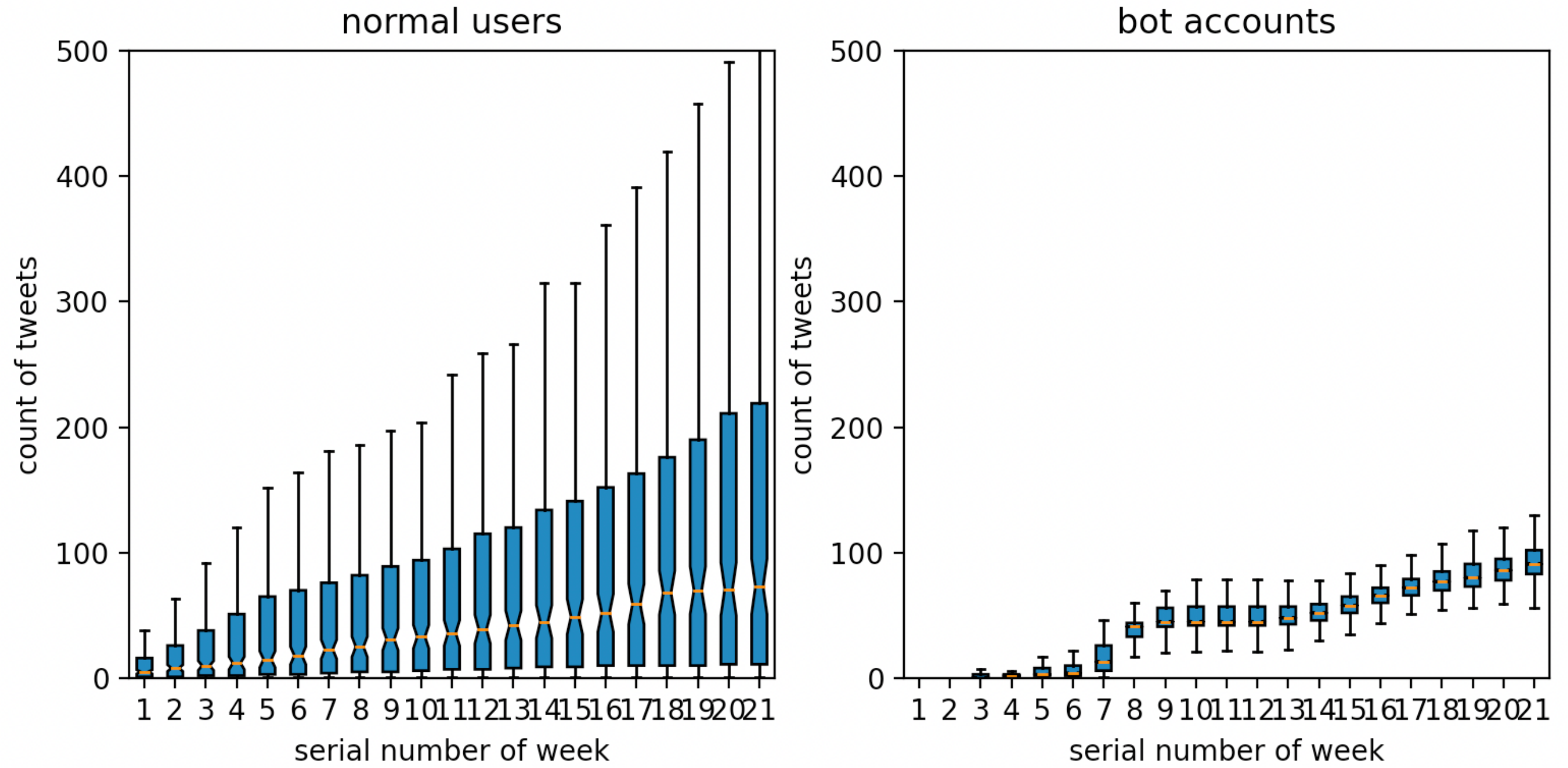}
\vspace{-3pt}\\
\mbox{\small (b) Weekly accumulated count of tweets}
\end{tabular}
\caption{Boxplots of tweeting counts: social bots vs normal users}
\label{fig:weekly_tweets}
\end{figure}

\begin{figure*}[h]
\centering
\begin{minipage}[h]{0.45\textwidth}
\centering
\includegraphics[width=3.0in]{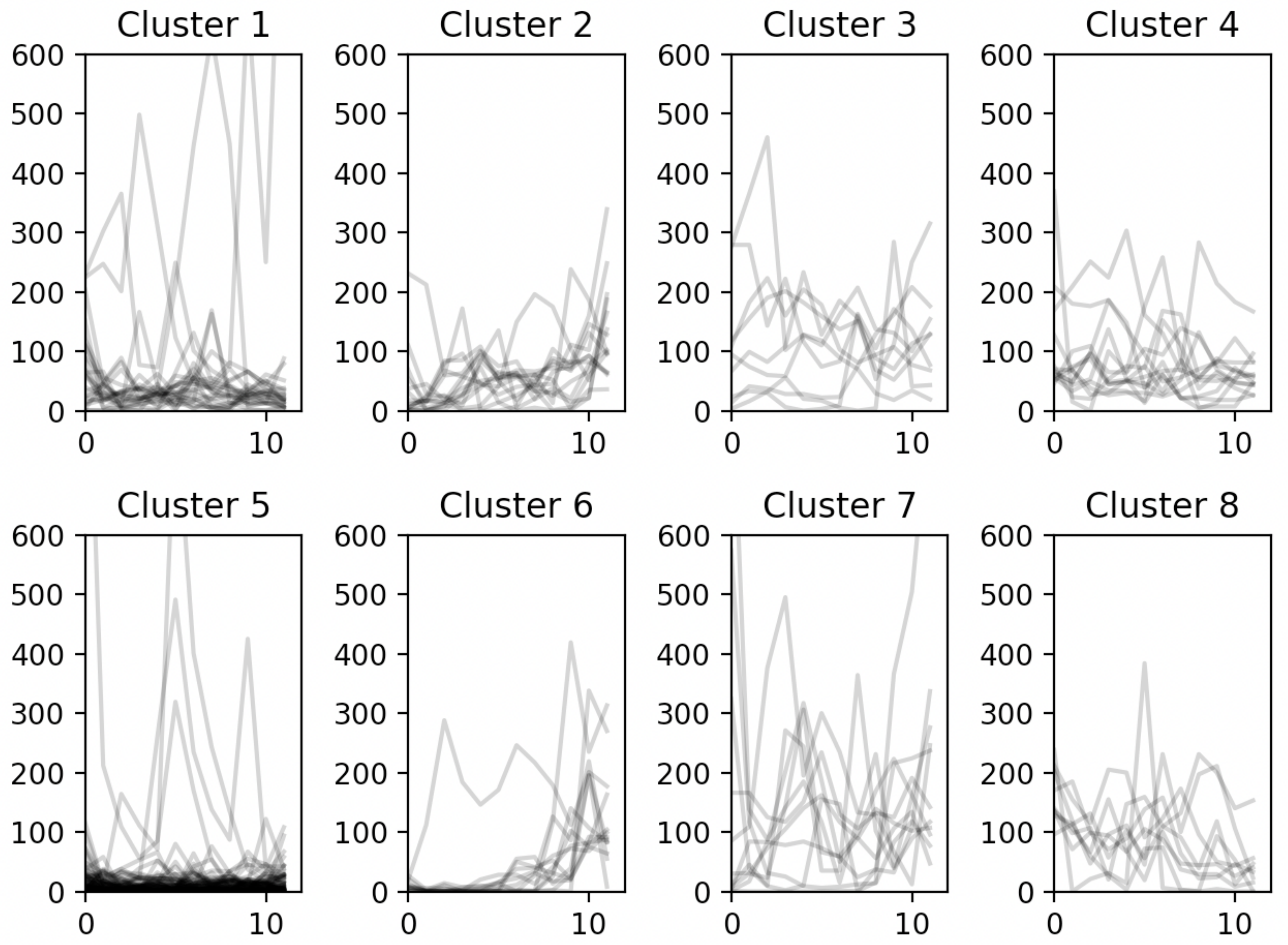}
\caption{Clusters of monthly behavioral sequence for normal users}
\end{minipage}
\hspace{0.5in}
\begin{minipage}[h]{0.45\textwidth}
\centering
\includegraphics[width=3.0in]{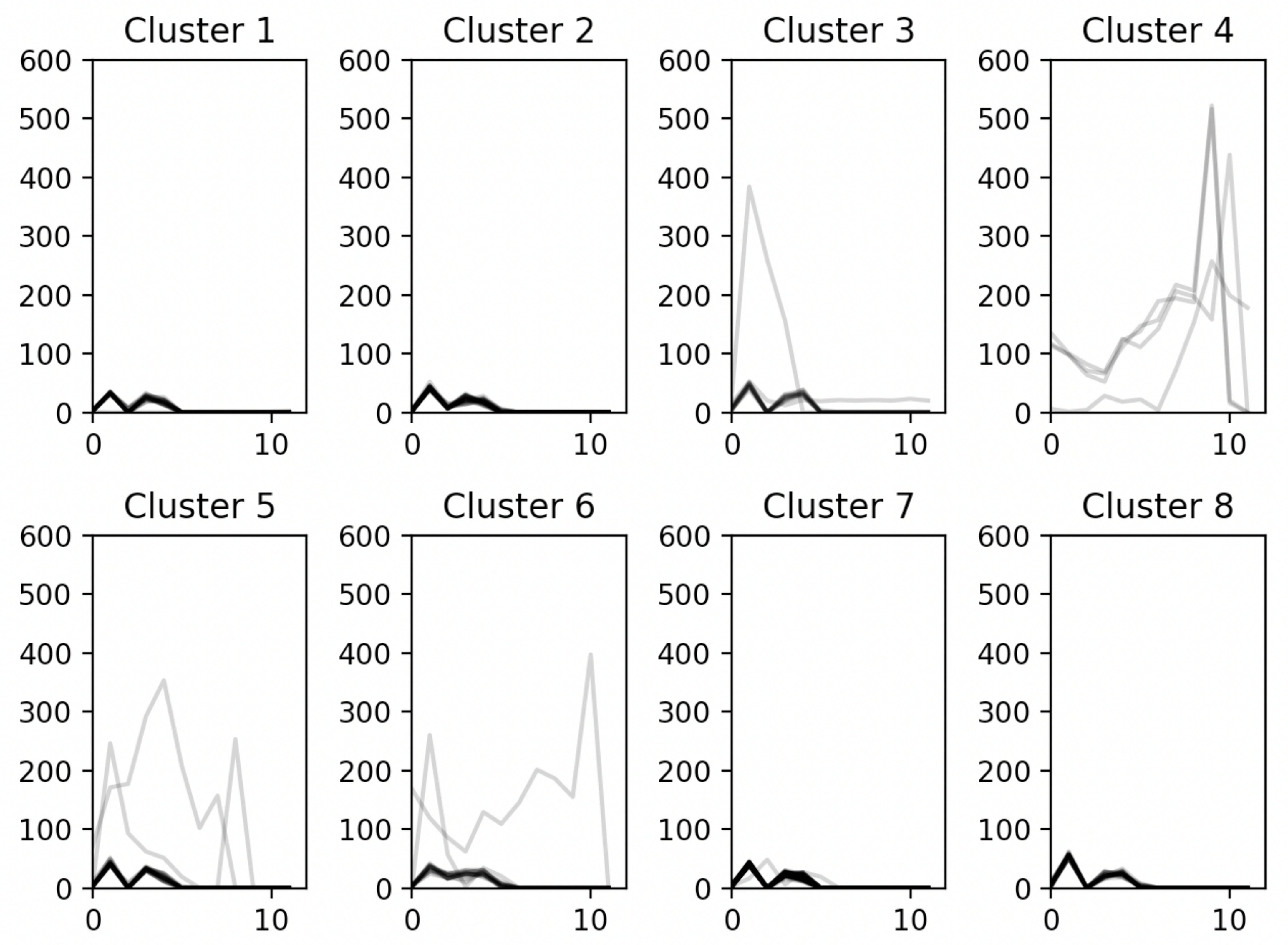}
\caption{Clusters of monthly behavioral sequence for social bots}
\end{minipage}
\label{fig:tweet_ts_cluster}
\end{figure*}

\subsection{Observe Behavioral Indicators Distribution in the Life-cycles of Accounts}

\subsubsection{\textbf{Behavioral statistic in a time-window}} 
The data set \Cresci has complete tweeting records with time stamps, contexts,  identifiers of accounts, and registering time stamps. Therefore, we set the analyzed indicator as the tweets count for each account in this section. We consider measuring tweeting behavior enough because the account independently controls tweeting behavior having no relationship with other accounts, unlike the count of followers and friends affected by other accounts. Therefore, it is the most direct and meaningful attribute to show accounts' behavior and capture attack traces.

To observe how one behavioral indicator fluctuates, we select various window sizes to calculate it. We first select one month as the time interval, meaning $dur$ equals 30, 60, 120, ... 360 (days). We also select one day as the time interval for a more fine-grain analysis, where $dur$ equals 1, 2, 3, ... 30 (days). $Indicator_{dur}$ equals the tweets count from the registration date to the $dur$ day.
We analyze how an indicator time series(called $Seq_{bhv}$) fluctuates across 12 months and 30 days. For a specific $dur$, we use box-plot \cite{potter2006methods} to present the distributions of $ indicator_{dur}$ consisting of all accounts. In the box-plot figure, each box shows the minimum value, first quartile, average, third quartile, and maximum value, which could clearly show the range and critical statistics of a bunch of data points.

\subsubsection{\textbf{A long-term observation}}

Figure~\ref{fig:monthly_tweets} shows the box-plot of monthly $Indicator_{dur}$. The upper left figure uses the data of normal users, and the upper right figure plots the data of social bot accounts. When comparing the two figures, we conclude that normal users are continually active, but bot accounts cooperate tacitly to reduce and even stop posting tweets after the sixth month. The following two figures are the accumulated tweet count from the registration to the corresponding month. The box plots of normal users show steady growth, but plots of bot accounts show a sudden termination starting in the fifth month.

All the results reflect each role of accounts. Most regular users have one account for one platform and persistently use it. In an online social occasion, changing into a new account tends to mean losing followers and friends, so regular users nearly do not change accounts. However, along with doing more and more spam, bot accounts will get more easily caught by the platform's risk control systems. What follows is that the cost of escaping detection and unlocking account improve quickly. As a result, the controller chooses to give up old bot accounts and register product new bot accounts for new spam purposes in the future.

\subsubsection{\textbf{The period at newly-registration}}
The monthly $dur$ interval is a bit coarse, in which account behavior patterns in newly-registered periods would hide. Therefore, we construct daily $dur$ indicators to watch the situation in the first month after registration. Figure~\ref{fig:weekly_tweets} is the box plot of tweeting count distribution at week-grain for three months (21 weeks). The two upper figures are the weekly count of tweets, and the two below are the accumulated count. The differences between bots and normal users are similar to monthly time granularity.

The normal users' tweeting box plot shows a burst of tweets in the first week, which means normal users often post more tweets at registration. However, bot accounts unexpectedly stay freezing (no tweeting behavior) until the third week after registration. Now look at the accumulation box plot: normal users show an apparent steady increase in tweet counts, but bot accounts have an unusual stopping of tweeting from the eighth week to the thirteenth week. In addition, the tweeting count of one normal user is larger than bot accounts on average.

\subsection{Mine Effective Patterns from Behavioral Sequences}

\subsubsection{\textbf{Busy and idle in a time-series seasonality}}
\label{subsec:busy_idle}

\begin{figure}[htbp]
\centering
\begin{tabular}{c}
\includegraphics[width=3.3in]{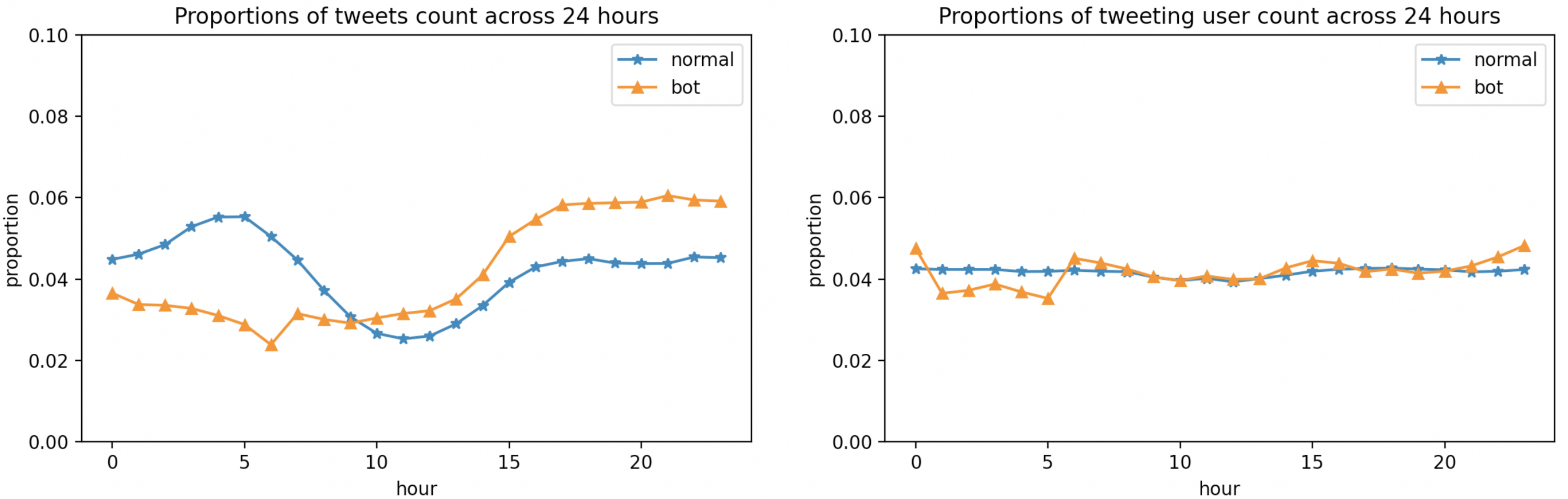}
\vspace{-3pt}\\
\mbox{\small (a) Percentages of tweets and tweeting users per hour}
\vspace{3pt}\\
\includegraphics[width=3.3in]{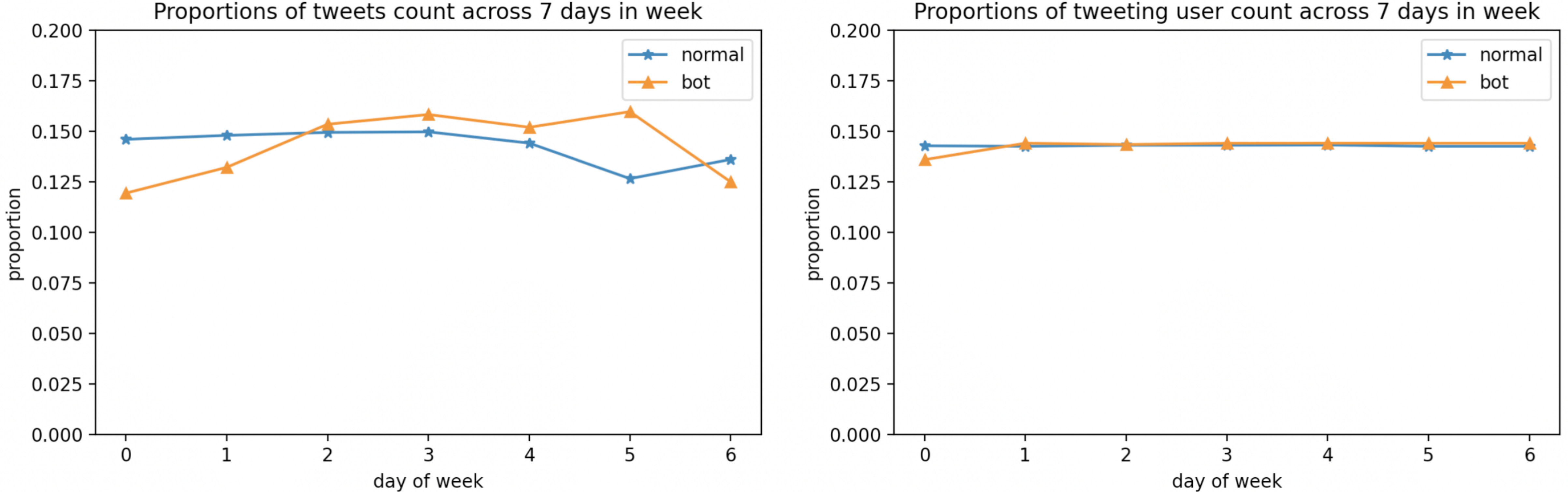}
\vspace{-3pt}\\
\mbox{\small (b) Percentages of tweets and tweeting users per day of week}
\end{tabular}
\caption{Decomposed seasonality from behavioral sequences of tweeting}
\label{fig:tweet_ts_count}
\end{figure}

Due to the regularity of using habits, user behavior statistics in social network platforms are often time series with seasonality. Also, the busy and idle time distribution are related to usage habits. For example, people sleep at night and then perform an idle period. We analyze the tweeting behavior time series properties based on a seasonal method. We group tweets into multiple parts according to time attributes, such as the hour of the day (from 0 am to 11 pm) and the day of the week (from Monday to Sunday).

Figure~\ref{fig:tweet_ts_count} shows the results. We use the proportions of the counts occupying the total amount rather than the actual tweeting counts. The peak and low of normal users and bots are different. Normal users peak at 5 am and low around 11 am. Bot accounts peak at 5 pm and low at 6 am. 

What is more different is that the trends of the two groups are nearly opposite, from 0 am to 9 am, which means normal users are active, but bots are inactive. The day-of-week distribution reflects a similar problem. We observe the conflict between peak and low and different local trends. Tweeting user count distributions in 24 hours and seven days are uniform and similar, whether normal users or bot accounts. It means the active proportions of accounts in the two groups are similar in each time unit.

\subsubsection{\textbf{Patterns of time series}} 
The wave's shape is critical for classifying and clustering time series. Several algorithms could extract shaping features, such as 
Dynamic Time Warping (DTW) \cite{berndt1994using}, Shapelets \cite{ye2009time} and deep learning method. \cite{bao2017deep}. We choose the DTW algorithm to cluster monthly tweeting count time series because it is highly efficient to discover similar time series patterns among multiple time series. We use tslearn \cite{tslearn} k-means and dynamic time warping library for computing. It uses DTW as core features in k-means algorithms and can group a large amount of 
time series into clusters quickly and precisely. 

For fairness, we randomly sample 200 tweeting behavior time series from the normal user and bot account groups, respectively. Figure~\ref{fig:tweet_ts_cluster} shows the result. It is obvious that the time series of bots are highly similar, reflected in that in each cluster, shapes of curves are similar, and overall outlines (think black lines in each cluster) of 5 clusters are similar. The clustering results of normal users are the opposite, with a phenomenon where in each cluster, curves are mussy, having no similarity without a clear overall outline. Also, there is nearly no similarity or correlation among the 6 clusters. We can conclude that bot accounts' behavior correlates and acts concurrently due to centralized control. Regular accounts, controlled by natural persons,  perform vastly different behavior patterns due to various usage habits.

All the indicators shown in measurement results could be transformed into account features for machine learning models to do classification. Also, the variances of those features between a bot and normal users could improve the separability of the two data points groups, making prediction more accurate.

\section{Design}

We present \BotShape, a social bots detection framework. It takes account logs as input, does an automatic behavioral feature-generating process, and then detects social bots based on machine learning classifiers. The feature engineering process has two steps: (i) extract behavioral sequences from account registration logs and event logs in various time intervals. (ii) compress raw behavior sequences time series to seasonality attributes and distinctive shapelets features \cite{ye2009time}. Behavioral features are actual multiple numerical vectors compatible with many machine learning classifiers. 

\subsection{Main Idea}

\subsubsection{\textbf{How to detect bots at an earlier time?}}
\begin{figure}[htb]
\centering
\includegraphics[width=\linewidth]{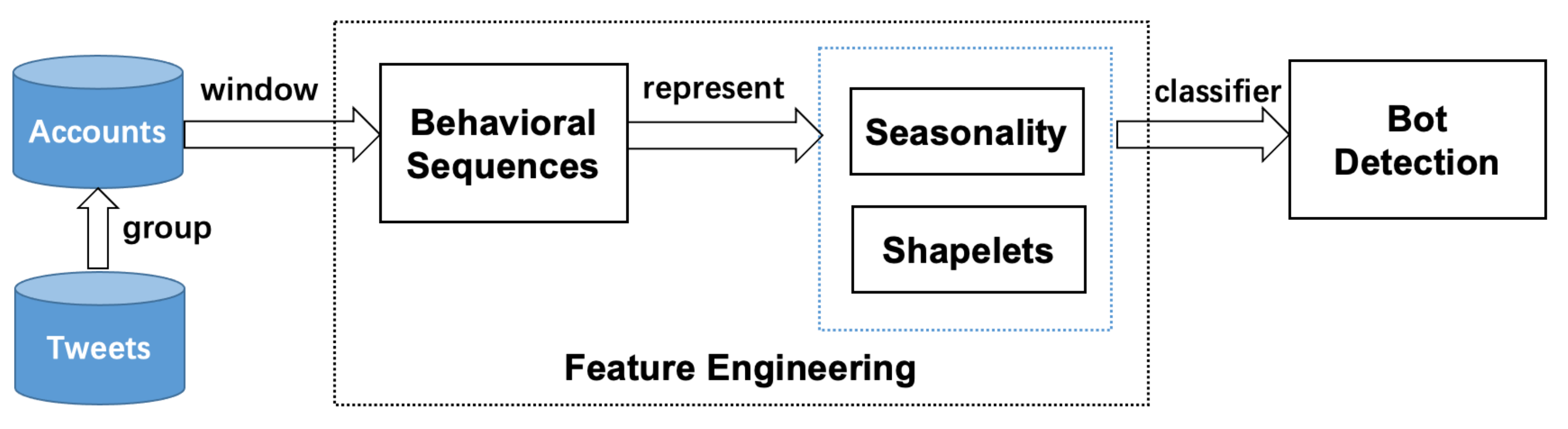}
\vspace{-6pt}
\caption{\BotShape architecture.}
\label{fig:arch}
\end{figure}

Figure~\ref{fig:arch} shows the \BotShape architecture, where each module is reasonable and has a corresponding supporting point in section~\ref{sec:measure}. \BotShape takes full advantage of the raw account event logs to discover the most compelling features for a classifier to distinguish bot accounts and user accounts. Different from \textbf{account-based} method, taking indicators accumulated for a more extended period as input, \BotShape could be deployed anytime after the registration due to its fine-grain observation view.

\subsubsection{\textbf{How to compress dimensionality and keep useful features?}} 
Extracting the best feature set is not trivial because the original features constructed by setting various time-related parameters are high-dimensional, leading to the \emph{Curse of Dimensionality} problem \cite{donoho2000high}. In other words, it will not give rise to a high detection accuracy if only piling up a mass of features. Therefore, it is necessary to mine prominent features from many raw behavioral series for the downriver model to make prediction. 

\subsubsection{\textbf{How to design an automatic system?}} We assess the automation degree of a  system based on the complexity of the manual parameter-setting process. Namely, it is highly automotive if a system could directly produce compelling features after setting a few parameters without too many attempts. We design \BotShape according to this principle. \BotShape applies a two-step separate feature engineering process with flexible parameter management, decoupling features election from raw feature production. The first step is generating a batch of statistical behavior sequences via different parameter settings such as time interval, period, and statistic functions. The second step focuses on mining the critical fluctuation points of those sequences based on the Shapelets method, which vastly reduces the dimensionality of feature vectors. Users could tune parameters respectively and flexibly for the two parts.

\label{sec:design}

\subsection{Behavior Sequence Features}

\label{subsec:design_bs}

\begin{algorithm}[t]
	\label{alg:bhv}
    \caption{Generate Behavioral Sequences (Time Series)}
    \begin{algorithmic}[1]
    \Require {\\ \textbf{Registration Logs:} $Log_{reg} = (id_{user}, t_{reg})$ \\
     \textbf{Event Logs:} $Log_{bhv} = \{(id, t_1), (id, t_2),..., (id, t_n)\}$ \\
     \textbf{Parameters:} $dur$, $gran$}
    \Ensure \emph{Behavioral Time-series} $ts$
    	
    \Function {$gen\_bhv\_sequence(Log_{bhv}, t_{reg}, dur, gran)$}{}
    	\State {$win = floor(\frac{dur}{gran})$}
    	\State {$ts =$ \textbf{new} $int[win]$}
    	\For {$k = 1, ....win $}
    		\State {$T_{st} = gran * (k-1)$}
    		\State {$T_{ed} = gran * k$}
    		\State {$cnt = 0$}
    		\For {$i=1, 2, ..., n$}
    			\If {$t_i - t_{reg} > T_{st}$ \textbf{and} $t_i - t_{reg} <= T_{ed}$} {$cnt = cnt + 1$}
    			\EndIf
    		\EndFor
    		\State {$ts[k] = cnt$}
    	\EndFor
    	\State{\textbf{return} $ts$}
    \EndFunction
    \end{algorithmic}
\end{algorithm}

We now elucidate the first step of the feature engineering process about generating behavior sequences using event logs $Log_{bhv}$ and registration information $Log_{reg}$. We format behavior sequences as multiple time series data. The program extracts behavioral time series from raw logs based on two time-related parameters. Parameter $dur$ is the duration from the registration to the computation time. Parameter $gran$ refers to the time granularity, such as day, week, and month, which is the window size for calculating statistical features like the number of tweets.

 Pseudo-code~\ref{alg:bhv} introduces how to generate behavioral sequences in a time series format. It takes two kinds of data as input, including the registration time stamp of an account and its event logs. Event logs could be any action on the social platform, like FOLLOW, SHARE, POST, and EDIT PROFILE. The computing is individual for each account, so this process can be deployed paralleled and distributed. In the generation process, it firstly computes the count of windows $win$, equaling $dur$ dividing $gran$, then it scans each time window to calculate the behavioral statistic like the count of tweets. In each round, the program sets the start time $T_{st}$ and end time of $T_{ed}$ each window for precisely distributing each $Log_{bhv}$ item to its time window. \BotShape outputs the sequences of behavioral statistics in chronological order, forming a time series. Engineers could generate a bunch of time series by setting various function parameters.
 
\subsection{Behavior Pattern Features}
\label{subsec:design_ts}
\subsubsection{\textbf{Seasonality Decomposition}}
In a time series data, \emph{seasonality} is the periodic fluctuation correlated strongly to the time attribute. Some time series data could perform a seasonal variation because the related entity changes regularly over time. For example, the number of online users in a network correlates with the working and sleeping time (busy or idle state) of people \cite{wu2018cellpad}, so the time series indicator could perform a repeat and similar pattern. Plus, social platforms for online activity are highly affected by user usage habits and tend to present hourly, daily, and weekly seasonality. Also, in sub section~\ref{subsec:busy_idle}, we compare the behavioral seasonality distributions of bots and genuine users, hourly and daily, respectively, showing an evident divergence.

Assuming that classifiers could separate bots from genuine users after exploring the seasonality of behavioral sequences, \BotShape owns a continued process of seasonality extraction. In detail, After generating a time series, \BotShape continues to compute the seasonality, which is still a sequence where each item is the mean value of corresponding elements. For example, for extracting the weekly seasonality extraction from a day-gran time series, \BotShape firstly queries the day of the week of each time-stamp and then allocates them to seven groups (Sunday, Monday, ..., Saturday), finally averages each group and organizes them into a new sequence in order.

\subsubsection{\textbf{Shapelets Representation}}
Shapelets \cite{ye2009time} are subsequences of a time series that are prominently distinct characteristics of its class. The Shapelets algorithm performs outstandingly in time series classification problems compared to raw data. The algorithm targets finding the best splitting strategy for maximizing the information gain (difference between entropy before and after the splitting). 

In sub section~\ref{subsec:busy_idle}, we observe an apparent similar time series shape in six clusters of bot behavioral sequences. If \BotShape could represent the pattern correctly, it would vastly improve the detection accuracy for bots. Therefore, \BotShape also takes continued Shapelet mining after constructing behavioral sequences. It applies a python package called tslearn \cite{tslearn} to learn shapelets. It learns from the raw time series and their labels from the train set to discover the best splitting points for the most accurate classification and then extracts the shapelets of the time series in the test set via the same splitting manner. In the default setting, \BotShape extracts shapelets from weekly and monthly behavioral sequences as new features, with a default subsequence length setting of 30\% of sequence length for weekly and 50\% for monthly.

\subsection{Bot Detection}
Behavioral data constructed by \BotShape, including behavioral sequence, seasonality, and shapelets, are potential features for machine learning classifiers to fit and predict. We emphasize that \BotShape focuses on behavioral feature engineering, providing extra general bot features, rather than \textbf{account-based} attributes. To evaluate its general utility, \BotShape integrates multiple prediction algorithms instead of fixing the classifier. We evaluate prediction accuracy in sub-section~\ref {subsec:eval_features} shows behavioral features all perform a very high detection accuracy when using different classifiers to predict. Eventually, secondary time series features (seasonality plus shapelets) perform better.

\section{Evaluation}
\label{sec:evaluation}

\subsection{Ground-truth and Metrics}

\textbf{Ground-truth} is information about known properties of some entities based on observation and expert knowledge. In the classification problem, it provides the precise class of each instance (called \textbf{label}). Ground truth plays a vital role in machine learning because it is a  benchmark for evaluating performance and finding the best from models with various algorithms, parameters, and features. The data set \Cresci has three types of bot accounts, so we reconstruct them into three ground-truth sets shown in Table~\ref{tab:gt}. Each set consists of bot accounts and genuine users, and we define bots as positive instances and genuine ones as negative instances. One set by one, we gradually escalate the scope of bots for a more detailed performance evaluation in three situations.

\begin{table}[h]
\centering
\caption{Ground-truth}
\label{tab:gt}
\vspace{-6pt}
\renewcommand{\arraystretch}{1.1}
\small
\begin{tabular}{|c|c|c|p{1.5in}|}
\hline
{\bf Set}  & {\bf Positive} & {\bf Negative} & {\bf Descriptions}\\
\hline
BotSet 1 & 4912 & 1083 &social bots, genuine users \\
\hline
BotSet 2 & 5912 & 1083 & social bots, traditional bots, genuine users \\
\hline
BotSet 3 & 9263 & 1083 & social bots, traditional bots, fake followers, genuine users \\
\hline
\end{tabular}
\end{table}

In a standard machine learning classification pipeline, engineers randomly split the instances of a ground truth into two parts, including \emph{train set} occupying 70 percent and \emph{test set} occupying 30 percent. \BotShape uses instances of the train set to fit the functional relationship between features and actual labels. Then it outputs predicted labels for instances in the test set after inputting features. Due to the difference between actual labels and predicted labels in the test set, there are four cases for each instance: (i) True Positive (TP): the actual and the predicted label are all positive; (ii) False Positive(FP): the actual label is negative, and the predicted is positive; (iii) True Negative (TN): the actual and the predicted label are all negative; (iv) False Negative (FN): the actual label is positive, and the predicted is negative.

We adopt two metrics \textbf{accuracy} and \textbf{f1-score} based on those four statistics. \textbf{Accuracy} equals $\frac{TP+TN}{TP+FP+TN+FN}$. It reflects the correct rate of all predictions, regardless of whether the actual class is positive. Generally, the higher the accuracy, the more right prediction occurs in an online system. \textbf{F1-score} combines two important metric \emph{precision} and \emph{recall}, equaling $\frac{2 \cdot precision \cdot recall}{precision+reall}$. \emph{Precision} is the rate of correct prediction in predicted positive instances, equaling $\frac{TP}{TP+FP}$. High precision refers to a lower false alarm rate, meaning fewer wrongly punished genuine users. \emph{Recall} is the bot discovery rate in actual positive instances, and it equals $\frac{TP}{TP+FN}$. Higher \emph{recall} stands for a less count of bots that the detection system will miss. \textbf{F1-score} could comprehensively reflects \emph{precision} and \emph{recall} performance, by averaging them. In general, the higher \emph{f1-score is}, the higher \emph{precision} and \emph{recall} are.

\subsection{Assess Effectiveness of Behavioral Features}
\label{subsec:eval_features}

We evaluate the bot detection performance of \BotShape three ground-truth. We split feature data into three groups: \textbf{account}, \textbf{behavior}, and \textbf{time-series}. \textbf{Account features} refer to user properties profiling an account, including five indicators directly coming from data set \Cresci: the count of statuses, the count of followers, the count of friends, the count of favorites, and the count of lists. Also, for fair competition, we add one indicator into the feature set: the total count of tweets within one year after registration. \textbf{Sequence features} (mentioned in sub section~\ref{subsec:design_bs}) consists of 3 types of raw behavioral sequences, including daily tweet count in the first 30 days, weekly tweet count in the first 53 weeks, and monthly tweet count in the first 12 months after registration. \textbf{Pattern features} (mentioned in sub section~\ref{subsec:design_ts}) have two significant features compressed from \textbf{behavior} feature set, including \emph{seasonality} and \emph{shapelets} features.

\begin{table*}[h]
  \centering
  \caption{Accuracy and F1-scores}
    \begin{tabular}{|c|c|c|c|c|c|c|c|c|c|c|c|c|c|}
    \toprule
    \multicolumn{2}{|c|}{\textbf{data set}} & \multicolumn{4}{c|}{\textbf{social bots}} & \multicolumn{4}{c|}{\textbf{social + traditional bots}} & \multicolumn{4}{c|}{\textbf{all bots + fake followers}} \\
    \midrule
    \multicolumn{2}{|c|}{\textbf{feature set}} & \textbf{account} & \textbf{sequence} & \textbf{pattern} & \textbf{gain} & \textbf{account} & \textbf{sequence} & \textbf{pattern} & \textbf{gain} & \textbf{account} & \textbf{sequence} & \textbf{pattern} & \textbf{gain} \\
    \midrule
    \multirow{2}[4]{*}{\textbf{SVM}} & \textbf{Accuracy} & 86.56\% & 96.82\% & \textbf{97.83\%} & \textcolor[rgb]{ .275,  .663,  .129}{\textbf{11.27\%}} & 80.94\% & 93.16\% & \textbf{95.71\%} & \textcolor[rgb]{ .275,  .663,  .129}{\textbf{14.77\%}} & 90.70\% & 96.32\% & \textbf{97.18\%} & \textcolor[rgb]{ .275,  .663,  .129}{\textbf{6.48\%}} \\
\cmidrule{2-14}          & \textbf{F1 Score} & 79.65\% & 94.45\% & \textbf{96.11\%} & \textcolor[rgb]{ .275,  .663,  .129}{\textbf{16.46\%}} & 74.44\% & 88.55\% & \textbf{91.90\%} & \textcolor[rgb]{ .275,  .663,  .129}{\textbf{17.46\%}} & 80.13\% & 90.63\% & \textbf{92.28\%} & \textcolor[rgb]{ .275,  .663,  .129}{\textbf{12.15\%}} \\
    \midrule
    \multirow{2}[4]{*}{\textbf{LR}} & \textbf{Accuracy} & 94.95\% & 86.46\% & \textbf{98.84\%} & \textcolor[rgb]{ .275,  .663,  .129}{\textbf{3.89\%}} & 94.41\% & 84.71\% & \textbf{97.27\%} & \textcolor[rgb]{ .275,  .663,  .129}{\textbf{2.86\%}} & 95.75\% & 89.54\% & \textbf{98.57\%} & \textcolor[rgb]{ .275,  .663,  .129}{\textbf{2.82\%}} \\
\cmidrule{2-14}          & \textbf{F1 Score} & 90.68\% & 64.90\% & \textbf{97.92\%} & \textcolor[rgb]{ .275,  .663,  .129}{\textbf{7.24\%}} & 88.13\% & 57.30\% & \textbf{94.90\%} & \textcolor[rgb]{ .275,  .663,  .129}{\textbf{6.77\%}} & 86.40\% & 58.77\% & \textbf{96.23\%} & \textcolor[rgb]{ .275,  .663,  .129}{\textbf{9.83\%}} \\
    \midrule
    \multirow{2}[4]{*}{\textbf{MLP}} & \textbf{Accuracy} & 82.97\% & 84.34\% & \textbf{98.94\%} & \textcolor[rgb]{ .275,  .663,  .129}{\textbf{15.97\%}} & 84.06\% & 84.06\% & \textbf{96.80\%} & \textcolor[rgb]{ .275,  .663,  .129}{\textbf{12.73\%}} & 89.54\% & 89.54\% & \textbf{98.93\%} & \textcolor[rgb]{ .275,  .663,  .129}{\textbf{9.39\%}} \\
\cmidrule{2-14}          & \textbf{F1 Score} & 45.35\% & 54.01\% & \textbf{98.11\%} & \textcolor[rgb]{ .275,  .663,  .129}{\textbf{52.76\%}} & 45.67\% & 45.67\% & \textbf{93.98\%} & \textcolor[rgb]{ .275,  .663,  .129}{\textbf{48.31\%}} & 47.24\% & 47.24\% & \textbf{97.17\%} & \textcolor[rgb]{ .275,  .663,  .129}{\textbf{49.93\%}} \\
    \midrule
    \multirow{2}[4]{*}{\textbf{RF}} & \textbf{Accuracy} & 97.22\% & 86.81\% & \textbf{98.69\%} & \textcolor[rgb]{ .275,  .663,  .129}{\textbf{1.47\%}} & 96.84\% & 97.62\% & \textbf{98.96\%} & \textcolor[rgb]{ .275,  .663,  .129}{\textbf{2.12\%}} & 98.19\% & 98.51\% & \textbf{99.17\%} & \textcolor[rgb]{ .275,  .663,  .129}{\textbf{0.98\%}} \\
\cmidrule{2-14}          & \textbf{F1 Score} & 94.97\% & 68.55\% & \textbf{97.68\%} & \textcolor[rgb]{ .275,  .663,  .129}{\textbf{2.71\%}} & 93.68\% & 95.44\% & \textbf{98.01\%} & \textcolor[rgb]{ .275,  .663,  .129}{\textbf{4.32\%}} & 94.96\% & 95.98\% & \textbf{97.73\%} & \textcolor[rgb]{ .275,  .663,  .129}{\textbf{2.76\%}} \\
    \bottomrule
    \end{tabular}%
  \label{tab:eval_clf}%
\end{table*}%

To prove that features constructed by \BotShape feature engineering approach (\textbf{behavior feature} and \textbf{time series feature}) have steady improvements on different classifiers, we select four widely used algorithms: Support Vector Machine \cite{hearst1998support} (SVM), Logistic Regression \cite{kleinbaum2002logistic} (LR), Multilayer Perceptron \cite{murtagh1991multilayer} (MLP) and Random Forest \cite{breiman2001random} (RF). We apply a famous machine learning python library scikit-learn \cite{pedregosa2011scikit} to implement all the classifiers.

Table~\ref{tab:eval_clf} shows the accuracy and f1-scores on all classifiers and feature groups. The result points out \textbf{pattern feature} performs best on two metrics across all classifiers, with all accuracy exceeding 97\% and a very high f1-score range from 92\% to 98\%. \textbf{Sequence feature} also presents high accuracy on SVM and RF, but it shows low f1-scores on classifier Logistic Regression and Multilayer Perceptron. We infer that irrelevant indicators in \textbf{behavior feature} become noise, making a simple classifier learn fake functional relationships between noisy features and labels. Random Forest and Support Vector Machine are more robust to noisy features.

\subsection{Compare with Other Approaches}
As sub-section~\ref{subsec:related} mentioned, there are three detection approaches: \textbf{accout-based}, \textbf{content-based}, and \textbf{graph-based}. Under the condition of \Cresci data set, we could contrast \BotShape with \textbf{account-based features}. We could not experiment with the \textbf{Graph-based} method because there are no edges like the following relationship in the data set. We do not implement \textbf{content-based} method because our approach has already achieved very high accuracy, and its used computation resource is very saving compared with content models.

In table~\ref{tab:eval_clf}, we define a new metric called \emph{performance gain} to measure the improvement of \BotShape, which is the difference between \emph{pattern features} and \emph{account features}. The value of \emph{performance gain} is influenced by the actual accuracy of \BotShape and the initial accuracy of \emph{account features} because the largest accuracy value is 100\%. F1-score is also the same. \emph{Gain of accuracy} ranges from 0.98\% to 11.27\%, with an average value of 6.53\%. \emph{Gain of f1-score} ranges from 2.71\% to 52.77\% with an average value of 19.23\%. 

\section{Conclusions}

We study the problem of detecting social bots in online social network platforms, inspired by novel account behavioral features mining and characterizing on a real-world dataset collected from \Twitter. We present \BotShape, an intelligent social bots detection framework consisting of three processes: account event logs pre-processing, behavioral sequences and patterns mining, and accurate bot prediction.
We target discovering novel but compelling fine-grain behavioral features which have never been systemically studied and evaluated by previous works. \BotShape first automatically generates behavior sequences, which are time series of statistics of different time-window, to profile various periods in one account's life-cycle. Based on apparent similarities of the characteristics and fluctuations of behavioral sequences after analyzing and clustering, our system second compresses raw sequences from high-dimensionality vector to low-dimensionality but more effective time series patterns based on the Shapelet algorithm. \BotShape was evaluated as generally accurate on three ground truths and four classification algorithms. Also, behavioral patterns as features also show an outstanding performance than \emph{account-based} features and raw behavioral sequences produced by the first process.

\bibliographystyle{unsrt}
\bibliography{paper}

\end{document}